\documentclass{article}

\usepackage{amsfonts}
\usepackage{amsmath}
\usepackage{amssymb}
\usepackage{mathrsfs}

\title{Universe creation on a computer}
\author{Gordon McCabe\footnote{e-mail: gordon.mccabe@tesco.net}}

\def\eqalign#1{\,\vcenter{\openup.7ex\mathsurround=0pt
 \ialign{\strut\hfil$\displaystyle{##}$&$\displaystyle{{}##}$\hfil
 \crcr#1\crcr}}\,}

\begin{document}

\maketitle

\begin{abstract}

The purpose of this paper is to provide an account of the
epistemology and metaphysics of universe creation on a computer.
The paper begins with F.J.Tipler's argument that our experience is
indistinguishable from the experience of someone embedded in a
perfect computer simulation of our own universe, hence we cannot
know whether or not we are part of such a computer program
ourselves. Tipler's argument is treated as a special case of
epistemological scepticism, in a similar vein to `brain-in-a-vat'
arguments. It is argued that Tipler's hypothesis that our universe
is a program running on a digital computer in another universe,
generates empirical predictions, and is therefore a falsifiable
hypothesis. The computer program hypothesis is also treated as a
hypothesis about what exists beyond the physical world, and is
compared with Kant's metaphysics of noumena. It is argued that if
our universe is a program running on a \emph{digital} computer,
then our universe must have compact spatial topology, and the
possibilities of observationally testing this prediction are
considered. The possibility of testing the computer program
hypothesis with the value of the density parameter $\Omega_0$ is
also analysed. The informational requirements for a computer to
represent a universe exactly and completely are considered.
Consequent doubt is thrown upon Tipler's claim that if a hierarchy
of computer universes exists, we would not be able to know which
`level of implementation' our universe exists at. It is then
argued that a digital computer simulation of a universe, or any
other physical system, does not provide a realisation of that
universe or system. It is argued that a digital computer
simulation of a physical system is not objectively related to that
physical system, and therefore cannot exist as anything else other
than a physical process occurring upon the components of the
computer. It is concluded that Tipler's sceptical hypothesis, and
a related hypothesis from Bostrom, cannot be true: it is
impossible that our own experience is indistinguishable from the
experience of somebody embedded in a digital computer simulation
because it is impossible for anybody to be embedded in a digital
computer simulation.\hfill \break


\end{abstract}

\section{The epistemology of universe creation on a computer}

F.J.Tipler has suggested that our universe could be a computer
program running on a computer in another universe, (see, for
example, p240-244 of Tipler 1989, and p206-209 of Tipler 1995).
Tipler imagines a perfect computer simulation of our universe,
which precisely matches the evolution in time of our own universe,
and precisely represents every property of every entity in our
universe. Such a simulation would simulate all the people who
exist in our own universe. Such simulated people, suggests Tipler,
would reflect upon the fact that they think, would interact with
their apparent environment, and would conclude that they exist.
Their experience would be indistinguishable from our own
experience, and Tipler infers from this that we ourselves cannot
know that we are not part of such a computer program. \textit{Ex
hypothesi}, there is nothing in our experience which could be
evidence that we are not part of such a program, hence, it might
be argued, we cannot know that we are not part of a computer
program.

This argument is a type of epistemological scepticism, similar to
Descartes' dreaming argument. Descartes raised the possibility
that one could experience a dream which is indistinguishable from
the experience of a conscious, waking individual. The sceptical
argument from this is that, \textit{ex hypothesi}, there is
nothing in one's experience which could be evidence that one is
not dreaming, hence one cannot know that one is not dreaming.

A modern version of this is the `brain in a vat' hypothesis.
Jonathan Dancy characterises this sceptical hypothesis as follows:
``You do not know that you are not a brain in a vat full of liquid
in a laboratory, and wired to a computer which is feeding you your
current experiences under the control of some ingenious
technician/scientist...For if you were such a brain, then,
provided that the scientist is successful, nothing in your
experience could possibly reveal that you were; for your
experience is \textit{ex hypothesi} identical with that of
something which is not a brain in a vat. Since you have only your
own experience to appeal to, and that experience is the same in
either situation, nothing can reveal to you which situation is the
actual one," (Dancy 1985, p10).

One can identify two distinct premises in this argument:

\begin{enumerate}
\def\theenumi{{\rm(\alph{enumi})}}
\item{It is possible for a brain in a vat to be fed experience
of an illusional world.}
\item{It is possible for that experience to
be indistinguishable from our own experience.}
\end{enumerate}

From these premises, the reasoning is as follows: Because the
experience of the illusional world would be indistinguishable from
one's own experience, it is not possible to know whether or not
one's own experience is experience of a real world, or experience
of an illusional world fed to a brain in a vat. Hence, it is not
possible to know whether or not one is a brain in a vat.

There is, however, a vital ambiguity in the argument. There are
two different senses in which real world experience could be
indistinguishable from illusional world experience. One could
claim either of the following two propositions:

\begin{enumerate}
\item{The experience of the illusional world would be indistinguishable
from the real world in terms of the detailed content of the
experience.}

\item{The experience of the illusional world would be
indistinguishable as experience from experience of the real world.
In other words, the form of the illusory experience would be
indistinguishable from the form of real-world experience.}
\end{enumerate}

It is not clear which of these claims Dancy is making. To
illustrate the differences between these claims, consider the
following scenarios:

Firstly, suppose that an individual is born in the real world,
grows-up in the real world, and experiences the real world for 30
years, developing a range of cognitive skills, and accumulating a
large collection of memories. Then, one night, whilst he lies
asleep, the individual is unknowingly drugged and kidnapped by a
scientist. As the victim lies unconscious in the scientist's
laboratory, his brain is removed and wired up to a computer. When
the individual is allowed to recover consciousness, he wakes up to
experience an illusional world controlled by the computer. Suppose
that the individual retains his memories of the real world. To
prevent the individual from having a reason to believe that he is
a brain in a vat, the experience of the illusional world must be
indistinguishable from the individual's experience of the real
world. Both the form and the detailed content of the individual's
illusory experience must be indistinguishable from his experience
of the real world. The illusional world must have the same spatial
layout and the same apparent history as that part of the real
world known to the victim, and the illusional world must evolve
according to the same laws that operate in the real world. The
victim must feel that he experiences his world, and influences
events in his world, with the same body that he possessed before
he fell asleep the previous night. The victim must not recognize
any difference between the real world and the illusional world
that is not explicable by the laws of the real world. The victim
must appear to perceive the same world he perceived before he fell
asleep the previous night.

If these conditions were satisfied, then the individual would have
no justification for believing that he is a brain in a vat. In
accordance with conventional definitions of knowledge, if the
individual would not be justified in believing that he is a brain
in a vat, then he could not know that he is a brain in a vat.

It is possible to imagine other sceptical scenarios which do not
require the detailed nature of the illusional world to be
indistinguishable from the detailed nature of the real world. If
an individual's memories of the real world are deleted or
suppressed, and apparent memories of an illusional world
completely different from the real world are added, then
experience of the illusional world would not give the individual
reason to believe that he experiences an illusional world. The
individual could experience an illusional world with a spatial
layout and history totally different to the spatial layout and
history of the real world. The illusional world could operate
according to laws different to those that operate in the real
world. Nevertheless, the experience of the illusional world would
be indistinguishable, as experience, from experience of the real
world. In other words, the form of the illusory experience, if not
the detailed content, would be indistinguishable from real-world
experience.

To take another example, if an individual were fed illusory
experiences from birth, that individual would have no memories of
the real world. Hence, experience of an illusory world completely
different from the real world in terms of detailed content, would
not give the individual reason to believe that he experiences an
illusional world.

It is not necessary to suppose that the individual who experiences
an illusional world is an unwilling participant. It is possible,
for example, that one's entire lifetime of experience upon
20th/21st century Earth, is part of a virtual reality game, played
on a distant planet in the far-future. The technology of the
far-future might enable game-players to play any role, in any
factual or fictitious world. The game technology might suppress
one's real-world memories, and supply the memories of the
character one is playing. If the game technology suppressed one's
real-world memories, one would be unaware of playing a virtual
reality game. The game technology might even suppress one's
real-world cognitive skills; one might experience birth, growth
and mental development in the game world. Either way, one would
have no memory of deciding to enter the game world. Once again,
the sceptical argument is that one's own experience is
indistinguishable from the experience of someone playing such a
virtual reality game, hence one cannot know whether or not one is
playing such a game.

Those sceptical arguments which require the detailed nature of the
illusional world to be indistinguishable from the detailed nature
of the real world, share a common point of vulnerability. It is
possible for the hypothesis supporting such sceptical arguments to
be false, and it is possible to know that it is false.

If the detailed content of the illusory experience is
indistinguishable from the detailed content of real experience,
then one can infer facts about the real world from one's
experience, irrespective of whether one's experience is illusory
or not. This allows one to determine, by scientific investigation,
whether the hypothesis which supports the sceptical argument, is
true or false.

For example, consider the brain in a vat argument. Recall that
this sceptical argument is based upon the premise that it is
possible for a brain in a vat to be fed experience of an
illusional world. Because the illusional world would be
indistinguishable, by hypothesis, from the real world, one's
sensory systems and neurophysiology in the illusional world would
be the same as one's sensory systems and neurophysiology in the
real world. Hence, one could learn about one's real world
physiology and neurology from one's experience, irrespective of
whether one's experience is experience of the real world, or the
illusional experience of a brain in a vat. One could not be led
into forming false beliefs about the kind of entity one is without
the violation of the indistinguishability condition.

Investigation of the human brain may reveal that it is impossible
for it to be stimulated in a way which would produce experience
indistinguishable from the experience of a person who is not a
brain in a vat. Thus, the hypothesis upon which the sceptical
argument is based, could be false. If one knew from
neurophysiology that it is not possible for one to be a brain in a
vat, then one would know that one is not a brain in a vat. When
Dancy characterises the brain-in-a-vat argument he states that
``you have only your own experience to appeal to," (Dancy 1985,
p10). This is false because one can also appeal to one's
scientific understanding, based upon both theory and empirical
evidence.

The other sceptical scenarios share this vulnerability:
neurophysiological investigation of the brain could reveal that it
is not possible for dreams to be indistinguishable from the
experiences of a waking individual; research in micro-electronics,
computer science, and human physiology, might conclude that
totally authentic virtual reality is not possible.

Those sceptical arguments which do not require the detailed nature
of the illusional world to be indistinguishable from the detailed
nature of the real world, are more robust. If the detailed nature
of the illusional world is different from the detailed nature of
the real world, then one cannot necessarily learn about real world
physiology and neurology from illusory experience. However, these
more robust sceptical scenarios are dependent upon the following
premise:

\begin{itemize}
\item{Either it is possible to delete or suppress an individual's
memories of the real world, and to replace them with apparent
memories of an illusional world, or it is possible to feed an
individual with illusory experience from birth.}
\end{itemize}

If this premise is false, then all the sceptical arguments which
concern illusional worlds might be refuted by empirical
investigation. It is, however, difficult to establish whether this
premise is true or false. If scientific investigation reveals that
it is impossible in our world to feed an individual illusory
experiences from birth, and that it is impossible in our world to
delete or suppress an individual's memories, and replace them with
apparent memories of an illusional world, then this alone does not
establish whether the premise is true or false. If our world is an
illusional world, and if the detailed nature of the illusional
world is different from the real world, then scientific
discoveries about our world, the illusional world, do not tell us
anything about the real world.

\hfill \break

It has been assumed in this section that it is possible to make a
distinction between the form and content of experience. If such a
distinction is not possible, then the sceptical scenarios must be
re-categorised as follows:

\begin{enumerate}
\item{An individual in our world experiences an illusional world
which is indistinguishable from experience of our world. The
individual is unaware that his experience is illusional precisely
because the illusional experience is indistinguishable from
experience of our world.} \item{An individual in our world
experiences an illusional world which is distinguishable from
experience of our world. The individual is unaware of the
difference, either because his memories of our world have been
deleted or suppressed, or because he has experienced the
illusional world from birth.}
\end{enumerate}

In case 1 the sceptical argument is as before, with the reference
to the content of experience omitted. In case 2, the sceptical
argument is as follows: If an individual in our world could
experience an illusional world which is distinguishable from
experience of our world, and if that individual could be made
unaware that what he experiences is illusional, then our own world
experience could be illusional experience, distinguishable from
the real world. We cannot know whether or not our experience is
experience of the real world, or experience of an illusional world
different from the real world.

\hfill \break

Tipler's computer program hypothesis differs in one respect from
the brain-in-a-vat type of hypothesis. The latter hypothesis
suggests that an individual in a real world could be fed
experiences of an illusional world, a world that does not
objectively exist. Tipler's computer program hypothesis suggests
that an entire universe could be created as a computer program,
and that many individuals could be created as part of the program.
This hypothesis does not merely suggest that there is a computer
program which is feeding illusory experiences to individuals who
exist in a real world. Instead, individuals capable of experience
are themselves created by the program, and the world they
experience is just as real relative to them, as our world is
relative to us. It is not Tipler's claim that we cannot know
whether or not our world is an illusional world. Instead, he
claims that ``we cannot know if the universe in which we find
ourselves is actually ultimate reality," (Tipler 1995, p208).
Tipler's claim is that we cannot know what level of reality we
experience; that we cannot know whether or not the universe we
experience has been created on a computer existing in another
universe.

However, the hypothesis that our own universe is indistinguishable
from a universe created on a computer, may be false. It will be
demonstrated in section 3 of this paper that physical predictions
follow from the hypothesis that our universe is a program running
on a \emph{digital} computer. For example, it follows that the
structure of the universe must be discrete, and that the spatial
universe must be compact. If these predictions are found to be
false, then it is impossible for our universe to be a program
running on a digital computer. If the predictions are falsified,
then our universe is distinguishable from a universe created on a
digital computer. Alternatively, if these predictions are found to
be true, then it remains possible for our universe to be a program
running on a digital computer. Empirical investigation is
necessary to determine if Tipler's computer program hypothesis is
possible.

Nick Bostrom has proposed a distinct computer program hypothesis
in which he proposes that future `posthuman' civilizations will
have the technological capability to create simulations ``that are
indistinguishable from physical reality for human minds in the
simulation," (Bostrom 2003, Section III). Bostrom's simulation
hypothesis is more anthropocentric than Tipler's hypothesis,
proposing not that an entire universe could be created as a
computer program, and not, as Tipler proposes, that every property
of every entity be simulated, but only ``whatever is required to
ensure that the simulated humans, interacting in normal human ways
with their simulated environment, don't notice any
irregularities," (ibid.). Bostrom proposes only that ``a posthuman
simulator would have enough computing power to keep track of the
detailed belief-states in all human brains at all times.
Therefore, when it saw that a human was about to make an
observation of the microscopic world, it could fill in sufficient
detail in the simulation in the appropriate domain on an as-needed
basis," (ibid.). With the possible exception of macroscopic
objects in inhabited areas, elements of the simulated world are
created on-demand for the purposes of perception, and are not, in
general, simulated independently of perception.

Whilst this paper does acknowledge the possibility of creating a
computer simulation in which merely the experience of the
participants in the simulation is indistinguishable from our own
experience, the paper concentrates on the possibility of creating
a universe on a computer which is indistinguishable from the
realist conception of our own universe, i.e. this paper
concentrates on the possibility of creating on a computer a
universe in which objects, properties and processes are simulated
independently of their perception by observers in the simulation.
Bostrom's hypothesis is less amenable to empirical test precisely
because it doesn't assume that empirical observation and
measurement is indicative of an independently existing world.
However, sections 4 and 5 of this paper, and, in particular, the
argument that a digital computer simulation of a system cannot
provide a realisation of such a system, carry equal weight against
the hypotheses of Tipler and Bostrom.

\section{The metaphysics of universe creation on a computer}

The hypothesis that our universe is a program running on a
computer in another universe is not merely a sceptical
epistemological hypothesis, but a metaphysical hypothesis, in the
sense defined below.

The term `metaphysics' seems to have at least two different
meanings. On the one hand, it is the study of that which possibly
exists beyond the physical world. On the other hand, it is a whole
group of philosophical subjects, such as the studies of time,
causation, substance, and universals. These subjects seem to be
united by the fact that they involve very general, foundational
study of the nature of things.\footnote{The historical reasons for
the double-meaning can be traced to Aristotle, as Barry Smith
explains: ``The books of Aristotle's Physics deal with material
entities. His Metaphysics (literally `what comes after the
Physics'), on the other hand, deals with what is beyond or behind
the physical world - with immaterial entities - and thus contains
theology as its most prominent part. At the same time, however,
Aristotle conceives this `metaphysics' as having as its subject
matter all beings, or rather being as such. Metaphysics is
accordingly identified also as `first philosophy', since it deals
with the most basic principles upon which all other sciences
rest," (Smith 1995, p373).}

For the purpose of this paper, metaphysics is defined to be the
study of that which possibly exists beyond the empirically
detectable world. In contrast, physics is defined to be the study
of the empirically detectable world. The hypothesis that our
universe is a program running on a computer in another universe,
is clearly a metaphysical hypothesis, in the very specific sense
defined here. The hypothesis is that the computer hardware on
which the program is running cannot be empirically detected by the
beings represented in the software, hence the hypothesis is
metaphysical rather than physical.

It is important to distinguish Tipler's hypothesis from a
metaphysically distinct proposal made by J.D.Barrow. Barrow
suggests that ``If we were to regard the Universe as a vast
computer...then we can readily envisage the laws of Nature as some
form of software which runs upon the particular forms of matter
that form the world of strings and elementary particles," (Barrow
1991, p160). In Tipler's computer program hypothesis, the computer
hardware is inaccessible to the people represented in the computer
program; the constituents of matter, elementary particles or not,
are just as much a part of the program software as the laws of
physics. Presumably, each different type of particle or field
would correspond to a different data type in the program. Each
individual particle or field would then correspond to an instance
of the relevant data type. In programming parlance, an instance of
a data type is called a data object. Hence, the constituents of
matter would correspond to data objects defined in the program.
The laws of physics would correspond to the algorithms which act
upon the data objects defined in the program. In general, entities
would correspond to data objects in a computer program, and
processes would correspond to algorithms. For example, an
individual electron would correspond to a data object, and the
Dirac equation would correspond to an algorithm capable of acting
upon any electron data object. To give another example, in the
geometrodynamical formulation of general relativity, a 3-manifold
$\Sigma$, and the tensor fields $(\gamma_i,K_i,\phi_i)$
representing the intrinsic geometry $\gamma_i$, extrinsic geometry
$K_i$, and matter fields $\phi_i$ at time $i$, would all
correspond to data objects. The geometrodynamical evolution
process would correspond to an algorithm which calculates
$(\gamma_{j+1},K_{j+1},\phi_{j+1})$ from $(\gamma_j,K_j,\phi_j)$.

After suggesting that our universe could be a computer program
running on a computer in another universe, Tipler goes one step
further, and claims that there is no need for a computer to be
running the program. The state of memory of a digital computer can
be treated as a long string of binary digits, and this represents
a natural number in binary notation. Given that a computer program
maps an initial memory state to a final memory state, a computer
program can be treated as a mapping on the set of natural numbers.
Tipler duly treats a program as an abstract mapping $\mathbb{N}
\rightarrow \mathbb{N}$, and claims that ``if time were to exist
globally, and if the most basic things in the physical universe
and the time steps between one instant and the next were
discrete," (Tipler 1995, p208), then our universe could be in
one-to-one correspondence with such an abstract object. Tipler
acknowledges that the most basic things in the physical universe
could be continuous, hence he proposes a further generalization of
what a simulation is: ``Let us say that a perfect simulation
exists if the physical universe can be put into one-to-one
correspondence with some mutually consistent subcollection of all
mathematical concepts," (ibid., p209).

This proposal does not merely suppose that mathematical Platonism
is true, that mathematical objects exist independently of the
physical universe, in an abstract realm. Nor does it merely
suppose that physical objects possess intrinsic mathematical
properties. Instead, it supposes that physical objects can be
identified with mathematical objects. As Barrow puts it, ``We
exist in the Platonic realm," (Barrow 1992, p282). Whilst this is
a fascinating idea, I shall restrict the discussion in this paper
to the hypothesis that our universe is a program running on a
computer in another universe.

The notion that there is something which exists beyond the
empirically detectable world has famous precedents in the history
of philosophy. Various types of thing have been postulated to
exist beyond the physical world: mental entities, theological
entities, and mathematical entities. These types of metaphysical
suggestion are of no relevance to this paper. Rather, the focus of
attention is the metaphysical hypothesis that there is something
non-mental, non-deistic, and non-mathematical, which exists beyond
our physical world. For example, Kant proposed that there are
things-in-themselves, so-called `noumena', which exist beyond the
empirically accessible world. The metaphysics of the computer
program hypothesis can be compared with the metaphysics of Kant's
noumena.

To recall, Kant suggested that there is a distinction between
noumena and phenomena. The noumena are things in themselves, and
the phenomena are the appearances of things in sensory perception.
There are three possible ways of defining noumena. The noumena
could be things which exist independently of sensory perception,
they could be things which exist independently of empirical
detectability, or they could be things which exist independently
of cognition altogether. Obviously, things which exist
independently of empirical detectability also exist independently
of sensory perception, and things which exist independently of
cognition also exist independently of empirical detectability.

If one merely stipulates that noumena are things which exist
independently of sensory perception, then noumena could simply be
things which are too small to see, like atoms and electrons.
Things which are too small to see are still empirically
detectable. As a classic example, an electron leaves a luminescent
trail in a Wilson Cloud Chamber. The electron is not directly
perceivable, but it is nevertheless detectable. Kant seems to
suggest that noumena exist independently of both sense perception
and empirical detectability of any kind. Further, Kant seems to
hold that noumena are beyond cognition altogether. The computer
program hypothesis holds that the states and processes of the
computer in another universe, exist beyond both sense perception
and empirical detectability, but these states and processes are
not beyond cognition. What exists beyond the physical world is
conceivable, according to the computer program hypothesis. In
contrast, Kant seems to hold that we cannot even conceive what
things in themselves are like.

Tipler's computer program hypothesis is consistent with a
threefold distinction between the phenomenal, the physical, and
the metaphysical. This corresponds to the distinction between
appearance, physical reality, and metaphysical reality.
Appearances and phenomena consist of sensory experiences such as
colours, sounds, and smells. Physical reality is the world
described by physics, the world of atoms, electrons, and
space-time. The hypothetical metaphysical reality consists of the
states and processes of a computer in another universe. In this
threefold distinction, space and time exist independently of
sensory appearances, whereas Kant believed that space and time are
merely the format into which sensory experience is arranged.
Unlike Kant, Tipler's proposal does not relegate space-time to the
merely phenomenal.

The computer program hypothesis is an interesting case because the
global metaphysics is drawn from local physics. The nature of what
lies beyond the entire physical universe (global metaphysics) is
drawn from the nature of the computer, a part of the physical
world (local physics).

\section{Deriving empirical predictions from
the metaphysical hypothesis}

This section proposes that Tipler's metaphysical hypothesis that
our universe is a program running on a digital computer, entails
that

\begin{itemize}
\item{The universe is discrete}
\item{The solutions to the fundamental evolution equations of physics must be
computable functions}
\item{The spatial universe has compact topology}
\end{itemize}

These predictions are empirically testable, hence Tipler's
metaphysical computer program hypothesis is empirically testable.
It will be demonstrated in this section that Tipler's computer
program hypothesis is potentially verifiable or falsifiable by
astronomical observation.\footnote{None of the predictions above
will be invalidated by the development of quantum computers.
Although quantum computers might be able to perform certain
calculations faster than computers based upon the notion of a
Turing machine, the collection of uncomputable functions for a
quantum computer is the same as the collection of uncomputable
functions for a Turing machine. Like existing computers, quantum
computers will possess a finite memory. And like existing digital
computers, a quantum computer will only be able to represent
discrete things.}

In Bostrom's computer simulation hypothesis it would be possible
for the simulators to create an `apparent' space-time in which the
universe appears to be continuous and of non-compact topology,
even though the simulated world is actually discrete and of
compact topology. The possibility of such illusions prevents
Bostrom's computer simulation hypothesis from having empirically
testable predictions. However, as mentioned at the end of the
opening section, Tipler hypothesizes a universe simulation which
creates every property of every entity, and does not countenance
the possibility of creating illusions for the participants in the
simulation. Because, in Tipler's hypothesis, empirical observation
and measurement is indicative of objects, properties and processes
which are simulated independently of their observation and
measurement, Tipler's hypothesis does have testable implications.

J.D.Barrow has claimed that if our universe is a computer program,
then all the laws of physics must involve computable functions,
(Barrow 1991, p205). A computable function is defined to be a
function whose value can always be calculated to arbitrary
precision by performing a finite sequence of well-defined steps,
often called an `effective procedure'.\footnote{i.e for any
function value $f(x)$ and error margin $\epsilon$, there is an
effective procedure which yields a rational number $r$ such that
$|f(x) - r| < \epsilon$.} Certainly, if a universe unfolds in time
on a computer, evolution equations must be used to calculate each
time-step from the preceding time-step, and a solution of those
evolution equations implemented on a computer must be a computable
function. If the solutions of the fundamental evolution equations
of physics were found to be non-computable functions, then the
computer program hypothesis would be falsified. Whilst the
computer program hypothesis therefore predicts that the solutions
to the fundamental evolution equations of physics must be
computable functions, computability would not be necessary to
represent, at once, an entire space-time on a computer.
Computability is only a requirement if the representation attempts
to calculate one aspect of the universe from another aspect. As
Tegmark remarks, ``since we can choose to picture our Universe not
as a 3D world where things happen, but as a 4D world that merely
is, there is no need for the computer to compute anything at all -
it could simply store all the 4D data," (Tegmark 1998, p26).

Note also that algorithmic compressibility is not a necessary
condition to represent a universe on a computer. A digital
representation of something is defined to be algorithmically
compressible if the length, in bits, of the shortest program
capable of generating that digital representation, is shorter than
the length, in bits, of the digital representation itself. Our
universe might not be algorithmically compressible, but might
still be digitally representable on a computer. What follows is an
attempt to derive more specific empirical predictions from
Tipler's computer program hypothesis.

To represent the entire universe on a computer one must use
either:

\begin{itemize}
\item{A unified theory of everything.}
\end{itemize}
or
\begin{itemize}
\item{A set of different theories, each with its own limited
domain of applicability, such that the set of domains covers the
entire universe.}
\end{itemize}

We do not, at present, have a unified theory of everything, but we
do have a set of different theories, which grow progressively
closer to covering the entire universe, in all its detail. Of
these, the only empirically verified theory which is capable of
describing the universe as a whole is general relativity. However,
although general relativity can represent the universe as a whole,
when it does so, it is only concerned with the large scale
structure of the universe. It cannot represent detail on all
length scales, as a unified theory of everything could be expected
to do. Nevertheless, because general relativity has been
empirically verified, the predictions of a unified theory of
everything would have to converge to the predictions of general
relativity within general relativity's domain of applicability.

The physical predictions derived from Tipler's metaphysical
computer program hypothesis will be derived from an examination of
how to represent a general relativistic universe on a computer.
This is perhaps a weak point of the strategy. The universe may not
be a 4-dimensional Lorentzian manifold, as it is represented to be
in general relativity. We do not know what type of thing a unified
theory of everything, incorporating a theory of quantum gravity,
would represent the universe to be. It is, therefore, a
provisional decision to consider a universe created on a computer
to be a general relativistic universe.

In addition, the predictions derived assume that a digital
computer is the only type of computer which has the potential to
simulate an entire universe. Although it isn't proven to be
impossible for an analog computer to simulate an entire universe,
the current evidence suggests that an analog computer cannot have
the representational capacity to do so. An analog computer uses
concrete (and continuous) physical quantities of one sort, (e.g.
electrical quantities or hydraulic quantities), to represent
concrete (and continuous) physical quantities of another sort,
(e.g. the varying height of tides). In other words, an analog
computer uses the concrete physical quantities of its physical
components to represent the physical quantities of the system to
be simulated. Early analog computers were constructed from levers,
cogs, cams, discs and gears, and used mechanical motions to
perform calculations. Modern analog computers tend to use
electrical quantities, such as voltage levels, to represent the
quantities of a simulated system, and specially designed circuits
are used to perform arithmetic upon these voltage levels. Whilst
an analog computer might use voltage levels to represent the
values of quantities on a simulated system, a digital computer
uses voltage levels to represent bits, and then sequences of bits
encode the values of quantities on a simulated system. Analog
computers tend to rely upon a mathematical resemblance between the
pattern of quantity-values possessed by the machine and the
pattern of quantity-values possessed by the simulated
system.\footnote{A good example of a mechanical analog computer is
an orrery, a clockwork device for simulating the solar system. The
actual position and motion of the balls representing the planets,
represents the actual position and motion of the planets.} Analog
computers do not use the versatile, encoded, abstract
representation of physical quantities that a digital computer
uses, and this limits their representational capacity.\footnote{As
an exception, Hava Siegelmann (1999) has proposed neural net
analog computers which are abstract encoders, like a digital
computer.}

It is often claimed that the variables of an analog computer are,
in fact, continuously variable, but this claim can be disputed.
Variables such as electrical voltage or fluid pressure are
probably discrete when they are reduced to the quantum level. Even
if there are other variables which are genuinely continuous, it
would still not be possible to precisely control their value.
Suppose for the sake of argument that voltage is continuously
variable. It would be impossible to set a precise input voltage
of, say, $5.34$V. The best one could ever do is to set an input
voltage within some interval, say $5.34\text{V} \pm 0.01$. This
point is better illustrated for the case of irrational numbers. It
is impossible to set an input voltage of $\pi$, and this is not
because of the limitations of current technology, but because
infinite precision is not attainable.

In general relativity, space is represented as a 3-dimensional
differential manifold, and space-time is represented as a
4-dimensional differential manifold. Whilst every manifold has the
cardinality of the continuum, a digital computer, as it is
currently understood, can only deal with discrete items of data.
The most crucial fact to recognize about a computer program is
that the data objects defined in it are built from $\mathbb{Z}$,
the set of integers. In contrast, the objects of analytic
mathematics are built from $\mathbb{R}$, the set of real numbers.
The memory of a classical, (i.e. non-quantum), digital computer
consists of electronic circuits which have two possible voltage
states. These voltage states are represented by binary digits,
otherwise known as `bits'. An element of memory is therefore
called a `bit'. Each bit of memory has two possible states,
represented as $0$ and $1$. The set of possible states of a bit
can be represented as $\mathbb{Z}_2 = \mathbb{Z}/2\mathbb{Z} =
\{0,1\}$, the additive group of integers, modulo 2. $\mathbb{Z}_2$
is a realisation of the cyclic group of two elements. Each byte of
memory, a string of $8$ bits, and the smallest addressable unit of
memory, can be represented by

$$(\mathbb{Z}_2)^8 = \mathbb{Z}_2 \times \mathbb{Z}_2 \times \mathbb{Z}_2
\times \mathbb{Z}_2 \times \mathbb{Z}_2 \times \mathbb{Z}_2 \times
\mathbb{Z}_2 \times \mathbb{Z}_2,$$ the 8-fold Cartesian product
of $\mathbb{Z}_2$. Thus, for a classical computer with $n$ bytes
of memory, the entire memory can be represented by
$(\mathbb{Z}_2)^{8n}$, a discrete mathematical structure of $8n$
dimensions. All the data objects defined in a program correspond
to regions of memory, hence the data objects defined in a program
are built from subsets of the discrete mathematical structure
$(\mathbb{Z}_2)^{8n}$.

The memory of a quantum computer consists of physical systems
which possess a quantum state space isomorphic to the
2-dimensional complex Hilbert space $\mathbb{C}^2$. Each such
memory element is referred to as a `qubit', or `Qbit'. A string of
$n$ qubits is represented by the $n$-fold tensor product of
$\mathbb{C}^2$. Hence, the state of 8 qubits is represented by a
vector in

$$
(\mathbb{C}^2)^8 = \mathbb{C}^2 \otimes \mathbb{C}^2 \otimes
\mathbb{C}^2 \otimes \mathbb{C}^2 \otimes \mathbb{C}^2 \otimes
\mathbb{C}^2 \otimes \mathbb{C}^2 \otimes \mathbb{C}^2
$$ As a consequence, the state of the $n$ qubits can be quantum
mechanically entangled.

Each qubit is considered to have a fixed basis, $\{v_0,v_1 \}$.
Each vector in the $n$-fold tensor product consists of a complex
linear combination of the $2^n$ basis vectors $\{v_{i_1} \otimes
\cdots \otimes v_{i_n}: i_1 = 0,1 ,..., i_n = 0,1  \}$. The
algorithms of a quantum computer correspond to unitary operators
upon these complex Hilbert spaces. Because $\mathbb{C}^2$
\emph{is} built from the set of real numbers, and because each
qubit $\mathbb{C}^2$ possesses a continuum of quantum states, it
might appear that a quantum computer can store an infinite amount
of information. This appearance, however, is deceptive. Whilst
there are a continuum of possible unitary operators on a qubit
Hilbert space, each quantum computer will only be engineered to
implement a finite collection. Moreover, each quantum computation
must cease with a measurement of the state of the $n$ qubits, and
this collapses the state from a linear combination of the basis
vectors into one of the fixed basis vectors, $v_{i_1} \otimes
\cdots \otimes v_{i_n}$. The initial state on which the unitary
transformations can operate is also such a state, and as Mermin
comments, ``the entire role of the state of the Qbits at any stage
of a succession of unitary transformations is to encapsulate the
probability of the outcomes, should the final measurement be made
at that stage of the process," (2002, p16). Thus, a quantum
computer, like a classical computer, possesses a finite number of
\emph{accessible} states. In fact, $n$ qubits of memory possess
exactly the same number of accessible states as $n$ bits of
memory, namely $2^n$. The data objects defined in a program
running on a quantum computer are discrete objects.

Because every manifold has the cardinality of the continuum, and
because digital computers can only represent discrete objects, it
is impossible to exactly represent a manifold on a digital
computer. It is, therefore, impossible on a digital computer to
exactly represent space and space-time as they are represented in
general relativity.

If space and space-time actually are manifolds, and if a manifold
cannot be exactly represented on a digital computer, then space
and space-time cannot be exactly represented on a digital
computer. If the space and space-time of our universe cannot be
exactly represented on a digital computer, then our universe
cannot be a computer program running on a digital computer in
another universe.

However, as already mentioned, the space and space-time of our
universe may not actually be manifolds. Space and space-time may
not exactly be as they are represented to be in general
relativity. Perhaps space and space-time are discrete, and perhaps
the manifolds of general relativity only provide an idealisation
of a discrete reality. The space and space-time of our universe
can only be exactly represented on a digital computer if space and
space-time actually are discrete.

Loop quantum gravity offers, perhaps, a mathematically rigorous
means to quantize general relativity, and loop quantum gravity
suggests that space is discrete in some sense. Using Ashtekar's
`new variables' approach, canonical general relativity can be cast
in the form of a canonical gauge theory, albeit a gauge theory
with additional constraints to the Gauss constraint. The
configuration space is a space of $SU(2)$-connections on a
principal fibre bundle over a 3-manifold $\Sigma$. In loop quantum
gravity, each closed curve (`loop') in the 3-manifold defines a
functional on the space of $SU(2)$-connections. This functional is
obtained by taking the holonomy of each connection around the
loop, representing that group element as an operator on a vector
space, and then taking the trace of that operator. Furthermore,
each `spin network' embedded in the 3-manifold defines a
functional on the space of $SU(2)$-connections. A spin network,
treated in isolation, is a discrete mathematical object consisting
of a graph, (a collection of vertices and edges), an irreducible
representation of $SU(2)$ assigned to each edge, and an
`intertwining' operator between such representations assigned to
each vertex. Such a graph embedded in the 3-manifold $\Sigma$
defines a functional on the space of $SU(2)$-connections by taking
the holonomy of a connection along each edge, using the
representations to obtain operators along each edge, forming the
tensor product of all those operators, tensoring that with all the
intertwining operators, and then contracting to obtain a number,
the number assigned to the connection, (Baez 1995, p19). Such
functionals turn out to be eigenvectors of operators which
purportedly represent the area of surfaces in the 3-manifold and
the volume of regions in the 3-manifold. Furthermore, these
operators have discrete spectra.

If one accepts that quantum theory provides a complete description
of a physical system, then, arguably, it is not the configurations
of the classical system which exist, but the quantum state
function(al). Hence, in the case of loop quantum gravity, the
3-manifold used to define the classical configuration space does
not exist. Rather, it is the state functional defined by the spin
network which exists.

Many important questions remain. For example, the dynamics of loop
quantum gravity remain intransigent, and there is no obvious
classical limit to the theory. Whilst it is claimed that area and
volume are discrete, what are they the area and volume of, if a
3-manifold does not exist? Are area and volume re-interpreted as
properties of spin networks?

The established means of finding a discrete approximation to a
manifold, is to find a cell complex which is homeomorphic to the
manifold. In particular, one tries to find a simplicial complex
which is homeomorphic to the manifold.\footnote{This should be
distinguished from the \emph{Whitehead triangulation} of a smooth
manifold, which is a functor rather than an isomorphism. In
dimension 6 and below, the equivalence classes of smooth manifolds
up to diffeomorphism are in one-to-one correspondence with the
equivalence classes of piecewise-linear (PL) manifolds up to
PL-isomorphism, and every PL-manifold is PL-isomorphic to a
simplicial complex, (Pfeiffer 2004, p17-18). This doesn't entail
that every smooth 4-manifold is such that its underlying
topological manifold is homeomorphic with a simplicial complex. In
dimension 4, there is a significant discrepancy between the
category of topological manifolds up to homeomorphism, and the
category of smooth manifolds up to diffeomorphism. There are
topological 4-manifolds which have no differential structure, and
therefore no Whitehead triangulation, and there are families of
smooth 4-manifolds which share the same underlying topological
manifold, up to homeomorphism, but which are pairwise
non-diffeomorphic, and therefore have distinct Whitehead
triangulations.} The schema of the simplicial complex is a
discrete mathematical object, which can be exactly represented on
a computer. By representing the schema on a computer, one
approximately represents the manifold.

If the schema of a simplicial complex is the natural discrete
approximation to a manifold, then, conversely, the manifold can be
said to be the natural continuum idealisation of the schema. If
space and space-time are actually discrete, but if they can also
be represented in a continuum idealisation as a 3-manifold and
4-manifold, respectively, then it is natural to suggest that space
is actually a 3-dimensional schema, and space-time is actually a
4-dimensional schema. Regge calculus is generally considered to be
the `discretized' version of general relativity, and Regge
calculus duly represents space and space-time as a simplicial
complex.

Loop quantum gravity demonstrates that, although space and
space-time might not be manifolds, they might not be the schema of
simplicial complexes either. However, if space and space-time
actually are discrete, it may be that they are best represented by
loop quantum gravity on small scales, and best represented by the
schema of simplicial complexes on large scales.

Some explanation of the mathematics is in order here. An n-cell is
an object which is homeomorphic with the n-ball in n-dimensional
Euclidean space, $\mathbb{D}^n = \{x \in \mathbb{R}^n : \| x \|
\leq 1\}$. For example, a 2-ball is a disc, bounded by a circle,
while a 3-ball is a solid ball bounded by a 2-sphere. Any polygon
is homeomorphic with a 2-ball, and is therefore a 2-cell. Any
solid polyhedron is homeomorphic with a 3-ball, and is therefore a
3-cell.

A cell-complex is obtained by pasting together any number of
cells, so that the faces of the cells are either disjoint, or so
that they coincide completely. A $3$-dimensional cell-complex is
obtained by pasting together $3$-cells in such a way that the
faces, edges and vertices of the cells are either disjoint, or
they coincide completely.

The most interesting type of cell is a simplex. A $0$-simplex is a
point, or `vertex'; a $1$-simplex is a line segment, or `edge'; a
$2$-simplex is a triangle; and a $3$-simplex is a solid
tetrahedron. By pasting together simplices, one obtains a
simplicial complex, (see Stillwell 1992, p23-24). A
$3$-dimensional simplicial complex is obtained by pasting together
solid tetrahedra. The schema of a $3$-dimensional simplicial
complex can be specified as follows. First, one declares all the
vertices in the complex. Next, one can specify which subsets of
the set of vertices correspond to simplexes. By specifying a pair
of vertices, $\{P_i,P_j\}$, one indicates that those vertices are
connected by an edge. One can then specify which triples
$\{P_i,P_j,P_k\}$ of vertices correspond to the faces, and finally
one can list which quadruples $\{P_i,P_j,P_k,P_l\}$ of vertices
correspond to the tetrahedra. One could alternatively give each
edge a name, and then specify which triples of adjoining edges are
connected by a face. One would then name each face, and specify
which quadruples of adjoining faces are connected by a
tetrahedron, (see Geroch and Hartle 1986, p546).

Although the manifold models of general relativity may be
idealisations, one particular manifold model may eventually be
verified by observation, to the exclusion of all others. To be
specific, either a Friedmann-Roberston-Walker (FRW) model, a small
perturbation of a FRW model, or an exact solution close to a FRW
model, may be verified by astronomical observation. If the
computer program hypothesis predicts that space or space-time is
actually the schema of a simplicial complex on large scales, then
the manifold model of the large-scale universe must be
homeomorphic with a simplicial complex whose schema can be
represented on a computer. It is therefore important to determine
which manifold models of general relativity can be discretely
represented on a digital computer by the schema of a simplicial
complex. If a particular manifold model were to be verified by
astronomical observation, but that model could not be represented
by a schema on a digital computer, then the hypothesis that our
universe is a computer program running on a digital computer would
be falsified.

Suppose, then, that one tries to represent space-time on a
computer with the schema of a 4-dimensional simplicial complex.
Unfortunately, it is not known if every 4-manifold is homeomorphic
to a simplicial complex, (Stillwell 1992, p247).\footnote{Not to
be confused with Markov's demonstration that the homeomorphism
problem is unsolvable for triangulated 4-manifolds. In other
words, given a pair of 4-manifolds which are homeomorphic to
simplicial complexes, there is no algorithm to determine if they
are mutually homeomorphic.} Hence, there may be 4-manifolds which
cannot be discretely represented by the schema of a simplicial
complex. If the space-time of the universe has a manifold
idealisation which does not have a homeomorphic simplicial
complex, then the space-time of the universe would not be
representable on a computer by the schema of a simplicial complex.
If there were no other means of discretely representing such a
4-manifold on a computer, then the space-time of the universe
would not be representable on a digital computer.

More seriously, because a computer can only store a finite amount
of data, it can only represent the schema of a finite simplicial
complex, a simplicial complex which contains a finite number of
simplexes. A finite simplicial complex can only be homeomorphic to
a compact manifold, hence only a compact 4-manifold is discretely
representable by a schema on a computer. Now, a compact
four-manifold can only possess a Lorentzian metric if its Euler
characteristic is zero. If one hypothesizes that the entire
4-dimensional history of our universe is a representation on a
computer, then one derives the potentially testable prediction
that the topology of our space-time must be compact, and of Euler
characteristic zero.

As an alternative, the geometrodynamical formulation of general
relativity employs a so-called `3+1' decomposition of space-time.
One chooses a 3-manifold $\Sigma$, and one studies the
time-evolution of the geometry and matter fields on $\Sigma$. As
the geometry and matter fields evolve, a 4-dimensional space-time
unfolds. Such a space-time will, of necessity, have the topology
of $\mathbb{R}^1 \times \Sigma$.

The geometrodynamical formulation is advantageous because of
Moise's triangulation theorem for 3-manifolds, (Stillwell 1992,
p25 and p242). Moise demonstrated that every 3-manifold is
homeomorphic with a simplicial complex; one says that every
3-manifold can be `triangulated'.\footnote{This coincides with the
Whitehead triangulation of a smooth 3-manifold. Every topological
manifold of dimension 3 or lower has a unique PL-structure, and
homeomorphic topological manifolds have PL-isomorphic
PL-structures, (Pfeiffer 2004, p37).} Although it is true that
every n-manifold can be triangulated for $n \leq 3$, it is, to
reiterate, unknown whether all 4-manifolds can be triangulated.

Moise's theorem means that any possible topology of the spatial
universe can be discretely represented with the schema of a
3-dimensional simplicial complex. Once again, however, a digital
computer can only represent the schema of a finite simplicial
complex. Whilst a compact 3-manifold is homeomorphic with a finite
simplicial complex, a non-compact 3-manifold can only be
homeomorphic with an infinite simplicial complex, a complex which
contains an infinite number of simplexes.

Only a compact 3-manifold can be homeomorphic with a 3-dimensional
simplicial complex whose schema is representable on a digital
computer. Hence, if our universe is a program running on a digital
computer, then our spatial universe must have a compact spatial
topology in a continuum idealisation. Tipler's hypothesis that our
universe is a program on a digital computer, predicts that the
spatial universe is discrete, and yields the potentially testable
prediction that our universe has compact spatial topology in a
continuum idealisation.

The prediction of compact spatial topology means that the
Euclidean $\mathbb{R}^3$ and hyperbolic $H^3$ FRW universes are
both inconsistent with Tipler's computer program hypothesis. The
only FRW universe which has both simply connected and compact
spatial topology, is the $S^3$-universe. Hence, the only simply
connected FRW universe which could be discretely represented on a
computer, is the $S^3$-universe. There are, however, a host of
multiply connected compact FRW universes. The spatial geometry of
each such universe is obtained as a quotient $\Sigma/\Gamma$ of a
simply connected Riemannian space form\footnote{A complete and
connected Riemannian manifold of constant sectional curvature is
called a Riemannian space form.} $\Sigma$, where $\Gamma$ is a
discrete, properly discontinuous, fixed-point free subgroup of the
isometry group $I(\Sigma)$, (O'Neill 1983, p243 and Boothby 1986,
p406, Theorem 6.5).

Compact FRW models exist for any value of sectional curvature k.
Of the 18 flat, $k = 0$, 3-dimensional Riemannian space forms, 10
are compact. Given that one can only create compact FRW universes
on a computer, it follows that one can only create 10
topologically different $k = 0$ FRW universes on a computer.

All of the 3-dimensional Riemannian space forms of constant
positive curvature are compact, hence they could all be created on
a computer.

Whilst there are compact and non-compact quotients $H^3/\Gamma$,
there are an infinite number of such compact quotients. The work
of Thurston demonstrates that `most' compact and orientable
$3$-manifolds can be equipped with a complete Riemannian metric
tensor of constant negative sectional curvature. This means that
`most' compact, orientable $3$-manifolds can be obtained as a
quotient $H^3/\Gamma$ of hyperbolic $3$-space.\footnote{The
meaning of `most' in this context involves Dehn surgery, (Besse
1987, p159-160).} One can therefore create an infinite number of
possible negative curvature FRW universes on a computer. However,
there is no compact $k = -1$ space form which is globally
homogeneous. $H^3$ itself is the only globally homogeneous
3-dimensional Riemannian space form of constant negative
curvature, and $H^3$ is, of course, non-compact. Given that one
can only create a compact universe on a computer, one cannot
create a $k = -1$ FRW universe on a computer which is globally
homogeneous. Thus, if our own universe is a globally homogeneous
$k = -1$ FRW universe, it cannot exist on a computer. However, a
locally homogeneous $k = -1$ FRW universe, with compact, multiply
connected topology, could exist on a computer, and it is only
\emph{local} homogeneity which our astronomical observations are
capable of detecting.

In practice it is difficult to test the prediction of compact
spatial topology. Observational evidence currently indicates that
our universe is a FRW universe, but there is no observable
parameter in a FRW model which determines the spatial topology.
Thus, there is no necessary link between the spatial topology of a
FRW universe and the value of the density parameter $\Omega_0$;
one cannot infer the spatial topology of our universe from
$\Omega_0$.

However, in a `small', compact, multiply connected universe, it is
possible to see around the entire universe. To understand this,
begin by recalling that a Riemannian manifold $(\Sigma,\gamma)$
has a natural metric space structure. The metric tensor $\gamma$
determines a Riemannian distance $d(p,q)$ between any pair of
points $p,q \in \Sigma$. The Riemannian distance $d(p,q)$ is
dimensionless, in the sense that it lacks any physical units. In a
FRW model, it is the scale factor $R(t)$ which introduces physical
units of distance. The physical distance between $p$ and $q$ at
time $t$ is $R(t)d(p,q)$. Because $R(t)$ has physical units, so
does $R(t)d(p,q)$.

For any FRW universe, one can calculate the maximum Riemannian
distance, $d_{max}$, that light has travelled by a time $t_0$,
which is considered to be the present time. The relevant equation
is

$$d_{max}(t_0) = \int^{t_0}_{0} \frac{c}{R(t)} \, dt$$

A civilization located at some point $p$ in space, will, at time
$t_0$, be able to see no further, in any direction, than a
Riemannian distance of $d_{max}(t_0)$. This distance can therefore
be referred to as the Riemannian horizon distance. It is, of
course, a dimensionless quantity.

Now, recalling that the diameter of a metric space is the supremum
of the distances which can separate pairs of points, it is a fact
that any compact Riemannian manifold is a metric space of finite
diameter. If one created, on a computer, a FRW universe in which
$(\Sigma,\gamma)$ were a compact Riemannian manifold of
sufficiently small diameter, $diam \,(\Sigma,\gamma)$, then the
Riemannian horizon distance $d_{max}(t_0)$ could exceed $diam \,
(\Sigma,\gamma)$ by the time $t_0 \sim 10^{10}$. If so, the
horizon would have disappeared for the observers in that universe.
They would be able to see their entire spatial universe. No point
of their universe could be separated from them by a Riemannian
distance greater than $diam \,(\Sigma,\gamma)$, so if
$d_{max}(t_0) \geq diam \,(\Sigma,\gamma)$, then they would be
able to receive light from all regions of their spatial universe.

In such universes, individual galaxies and clusters of galaxies
would produce multiple images upon the celestial sphere of
planet-bound observers, (see Ellis 1971). Different compact
spatial topologies and geometries would produce different patterns
of ghost images and multiple images upon the celestial sphere.

However, although compact spatial topology is a necessary
condition for the entire spatial universe to be visible, it is not
a sufficient condition. Our universe might have compact spatial
topology, but if it is a `large' compact universe, then all of
space will not be visible. For all of space to be visible when the
universe is only $\sim 10^{10}$ yrs old, the Riemannian manifold
$(\Sigma,\gamma)$ which represents the spatial universe must be
sufficiently small, as well as compact, Even if our spatial
universe is small and compact, it would be extremely difficult to
identify multiple images of galaxy clusters. Hence, although the
presence of multiple images would verify the hypothesis of a
small, compact universe, the fact that they have not been
identified at the current time does not falsify the hypothesis. A
better means of testing the hypothesis is to search for paired
circles in the microwave background radiation. Recent research
indicates that if such paired circles exist, then one could derive
the spatial topology from the specific pattern of paired circles,
(see Cornish, Spergel, Starkman 1998). The CMBR power spectrum can
also be used to determine whether our spatial universe is a small
compact universe. A small compact universe would affect the CMBR
power spectrum on large angular scales. The WMAP satellite has
revealed anomalies in the CMBR power spectrum on large angular
scales. The quadrupole $l=2$ mode was found to be about $1/7$ the
strength predicted for an infinite flat universe, while the
octopole $l=3$ mode was $72\%$ of the strength predicted for such
a non-compact $k=0$ universe, (Luminet \textit{et al} 2003, p3).

The presence of paired circles or specific anomalies in the CMBR
power spectrum would verify that the universe is spatially
compact, and would thereby verify Tipler's computer program
hypothesis. Unfortunately, the absence of paired circles or
anomalies in the power spectrum would not entail that the spatial
universe is non-compact. Our universe could simply be a large
compact universe. Hence, the absence of paired circles or
anomalies in the power spectrum would not falsify the computer
program hypothesis.

Predictions about the lifetime of our universe are easier to test
than predictions about the spatial topology. The lifetime of our
universe is determined by parameters such as the Hubble parameter
$H_0$ and the density parameter $\Omega_0$, which can be inferred
from observation. Hence, if the computer program hypothesis made
predictions about the lifetime of our universe, it would be easier
to test it. If a universe is represented by a Lorentzian manifold
$(M,g)$, then the lifetime of the universe corresponds to the
`timelike diameter' of $(M,g)$. The timelike diameter of $(M,g)$
is the supremum of the length of all past-directed timelike curves
in $(M,g)$. As Beem and Ehrlich comment, ``the timelike diameter
represents the supremum of possible proper times any particle
could possibly experience in the given space-time," (Beem and
Ehrlich 1980, p329).

If a Lorentzian manifold with an infinite timelike diameter were
represented by a numerical solution of the Einstein
geometrodynamical equations, and if the size of the time steps in
the numerical solution were constant, then an infinite number of
time steps would be necessary. An infinite amount of information
would have to be processed. The ever-expanding $k \leq 0$ FRW
universes are examples of universes with an infinite lifetime. If
a computer in a universe with an infinite lifetime could process
information at a constant rate, then it could process an infinite
amount of information. However, an ever-expanding universe will
suffer an entropy `heat death', the amount of free energy
available converging to zero as $t \rightarrow \infty$.
Brillouin's inequality entails that there is a minimum, positive
amount of free energy which must be expended to irreversibly
process, or erase, a bit of information. Where $\Delta I$ is the
amount of information processed in bits,

$$\Delta I \leq \Delta E/k_B T \ln 2 \,.$$ $\Delta E$ is the free
energy expended, $T$ is the absolute temperature in degrees K, and
$k_B$ is Boltzmann's constant, (Barrow and Tipler 1986, p661). At
first sight, this suggests that it is impossible to process an
infinite amount of information in an ever-expanding universe
because the amount of free energy converges to zero. However, the
amount of energy which must be expended per bit of information
processed is temperature dependent. From the inequality above, one
can derive the following constraint on the rate at which
information can be processed:

$$
dI/dt \leq  \frac{dE/dt}  {k_B T \ln 2} \, .
$$ In turn, this entails the following constraint on the total
amount of information $I$ which can be processed between the
current time $t_0$ and some future time $t_f$, which might be
$\infty$:

$$
I = \int_{t_0}^{t_f} \frac{dI}{dt} dt \leq (k_B \ln 2)^{-1}
\int_{t_0}^{t_f} T^{-1} \frac{dE}{dt} dt \, .
$$

If the temperature converges to zero, $T \rightarrow 0$, as it
does in an ever-expanding $\Omega_0 \leq 1$ universe, then the
amount of free energy which needs to be expended per bit converges
to zero. Hence, although the amount of free energy converges to
zero, so also does the amount of free energy which needs to be
expended per bit. Thus, because the integral $\int_{t_0}^{\infty}
T^{-1} (dE/dt) \, dt$ can diverge, it may still be possible to
process an infinite amount of information in an ever-expanding
universe, even if the total free energy expended
$\int_{t_0}^{\infty} (dE/dt) \, dt$ is finite, (Barrow and Tipler
1986, p663).

In $\Omega_0 > 1$ universes, because the temperature diverges near
the final singularity, the rate at which free energy is expended
$dE/dt$, and therefore the total energy expended $\int_{t_0}^{t_f}
(dE/dt) \, dt$, must diverge if the total information processed is
to diverge.

If $\Omega_0 \leq 1$ in our universe, as current astronomical
evidence indicates, then our universe has an infinite timelike
diameter. Assuming that the simulation of such a universe would
require an infinite amount of information to be processed, the
possibility of the computer program hypothesis then rests upon
whether it is physically possible for the integral
$\int_{t_0}^{t_f} T^{-1} (dE/dt) \, dt$ to diverge in either a
$\Omega_0 \leq 1$ universe or a $\Omega_0 > 1$ universe. In both
cases this remains a matter of debate. If it is not physically
possible for the integral to diverge in either case, and if the
observation that $\Omega_0 \leq 1$ in our universe is reliable,
then could one conclude that our universe is not a program running
on a computer in another universe? If it is impossible to process
an infinite amount of information, then the only type of universe
which could be \emph{entirely} simulated on a computer would be a
finite lifetime universe. However, as Bostrom might perhaps
suggest, it remains possible that a partial simulation of a
$\Omega_0 \leq 1$ universe could be created on a computer in
another universe, a finite lifetime subset of the entire $\Omega_0
\leq 1$ universe. Hence, even if our universe is a $\Omega_0 \leq
1$ universe, and even if it is impossible to process an infinite
amount of information, our universe could be a finite lifetime
simulation running on a computer in another universe.

\hfill \break

Not only could Tipler's computer program hypothesis be falsified
by empirical investigation, as considered above, but there are
logical constraints upon what it is possible to simulate on a
computer.

A computer is a finite volume subsystem of a universe, which is
capable of representing the state of other systems. A system can
represent, exactly and completely, the state of another system, if
and only if the amount of information which can be coded in the
first system is greater than or equal to the amount of information
which can be coded in the other system. An entire universe is a
special type of system. Hence, a subsystem of a universe
$\mathscr{A}$ can represent, exactly and completely, the state of
a universe $\mathscr{B}$, if and only if the amount of information
which can be coded in the subsystem of $\mathscr{A}$ is greater
than or equal to the amount of information which can be coded in
universe $\mathscr{B}$.

As a special case, if the amount of information which can be coded
in a subsystem of a universe $\mathscr{A}$ is less than the amount
of information which can be coded in the entire universe
$\mathscr{A}$, then it is impossible for the subsystem of universe
$\mathscr{A}$ to represent, exactly and completely, the entire
universe $\mathscr{A}$.

The amount of information which can be coded in a system is
determined by the number of different possible states of the
system. If $N$ denotes the number of possible states, then the
amount of information $I$ which can be coded, in bits, is simply
$I = log_2 N$. Hence, if the number of possible states of a
subsystem of a universe is less than the number of possible states
of the entire universe, then it is impossible for that subsystem
to represent, exactly and completely, the entire universe.

However, just because a system is a subsystem of a universe, it
does not follow that the number of possible states of the system
is less than the number of possible states of the universe. True,
if the number of possible states of a subsystem is finite, then by
virtue of being a subsystem, that finite number must be smaller
than the number of possible states of the entire universe. For
every state of a subsystem, there must be multiple states of the
entire universe which induce the same state upon that subsystem,
hence the number of possible states of the entire universe must be
larger. However, if the number of possible states of a subsystem
is not finite, then it is possible that it has the same number of
possible states as the entire universe. \textit{A priori}, it is
quite possible that a subsystem of a universe $\mathscr{A}$, and
the entire universe $\mathscr{A}$, both possess an infinite number
of states. If the state space of a subsystem has the same
cardinality as the state space of the entire universe, then, by
definition, there exists at least one bijective mapping between
the two state spaces. Any such bijective mapping would enable the
states of the entire universe to be represented by the states of
the subsystem.

This argument can be presented in another way. If a subsystem
$\mathscr{S}$ of our universe represents the entire universe
$\mathscr{U}$, it must also represent $\mathscr{S}$ representing
$\mathscr{U}$. If it does this, it must also represent
$\mathscr{S}$ representing $\mathscr{S}$ representing
$\mathscr{U}$. And so on, \textit{ad infinitum}. This is possible
only if the subsystem can store an infinite amount of information.

If the entire universe only has a finite number of possible
states, then a subsystem will also have a finite number of states,
and the number of subsystem states will be smaller than the number
of universe states. However, if the entire universe has an
infinite number of possible states, then it is possible for a
subsystem to have either a finite number or an infinite number of
possible states.

If the entire universe has an infinite number of possible states,
then it could conceivably possess either a continuous infinity of
possible states, or a discrete infinity. A digital computer could
only represent the universe exactly if the universe is discrete,
hence the only case of interest here is the case in which the
universe has a discrete infinity of possible states. A digital
computer could only represent the universe exactly and completely
if the entire universe and the computer subsystem of the universe
both possess a discrete infinity of possible states. In other
words, a digital computer could only represent the universe
exactly and completely if both the computer and the universe can
code a discrete infinity of information.

A computer is a finite volume subsystem of the universe, hence to
determine if a computer could code the same amount of information
as the entire universe, it is necessary to determine if a finite
volume subsystem can code a finite or infinite amount of
information. To answer this question, it is necessary to determine
what the physical structure of the universe is.

At present, it appears that there are discrete levels of physical
structure in the universe. All macroscopic material objects in our
universe are composed of chemical elements and chemical compounds.
The latter are composed of atoms in different combinations and
organizations. Atoms are composed of electrons and atomic nuclei.
The nuclei of atoms are themselves composed of protons and
neutrons, which are themselves composed of quarks. The parts of
material objects do not appear to lie on a continuum.

Electrons and quarks are purported to be elementary particles,
pieces of matter which have no parts. If elementary particles do
exist, then our universe could be said to have a finite lower
level of structure. There would be no levels of structure below
the level of elementary particles.

I propose that a finite volume subsystem is limited to coding a
finite amount of information if and only if the following three
conditions are satisfied:

\begin{itemize}
\item{The number of structure levels available in a finite volume of
space is finite.}
\item{On each structure level, there is a finite set of parts in a
finite volume of space.}
\item{Each of the parts on each level of structure
has a finite set of states.}
\end{itemize}

A finite volume subsystem which satisfies these conditions has
only a finite number of possible states, and therefore cannot code
the same amount of information which can be coded in the entire
universe.

To reiterate, a computer could only represent the universe exactly
and completely if a finite volume subsystem can code a discrete
infinity of information. It seems safe to assume that, on each
level of structure, there is a finite set of parts in any finite
volume of space. The Bekenstein bound\footnote{Otherwise known as
the universal entropy bound.} and the so-called holographic bound
of Susskind and 't Hooft, purportedly entail that the parts on
each level of structure have a finite set of states, (Bekenstein
2003). Moreover, the existence of elementary particles would mean
that there is a finite set of structure levels in each finite
volume of space. It would appear, therefore, at first sight, that
all three conditions are satisfied. It would appear that a finite
volume subsystem cannot code a discrete infinity of information,
and it would appear that a computer cannot represent the universe
exactly and completely.

However, further thought raises some doubts. Both the Bekenstein
bound and the holographic bound place an upper limit on the
\emph{entropy} within a finite volume of space. Given a finite
quantity of weakly self-gravitating energy $E$ in a spherical
volume of radius $R$, which is isolated from other systems,
(Bekenstein 2004), the entropy $S$ is subject to the following
upper bound:

$$
S \leq 2\pi E R/\hbar c \; .
$$

The holographic bound is independent of the quantity of energy,
and places the following limit on the entropy of a spherical
volume of radius $R$, which is isolated from other systems:

$$
S \leq \pi c^3R^2/\hbar G \; .
$$

In both cases, it is then assumed that a finite upper limit to the
entropy of a finite volume of space entails a finite upper limit
to the information storage capacity of that volume. This might be
inferred from the following relationship:

$$\text{Information} = \text{Maximum entropy} - \text{entropy} \; .$$

By implication, it is the \emph{statistical states} or
\emph{macrostates} of a system which are the bearers of entropy
and information here. The states which provide a complete,
detailed description of a system are referred to as `microstates'.
A statistical state expresses only partial knowledge of the state
of a system, and, in classical mechanics at least, corresponds to
a probability distribution $\rho$ defined upon the space of
microstates $\Gamma$. A macrostate is a set of macroscopically
indistinguishable microstates $\Gamma_M \subset \Gamma$, and
corresponds to a special type of statistical state in which the
probability distribution is of a constant value $|\Gamma_M|^{-1}$
on $\Gamma_M$, and zero elsewhere.\footnote{$|\Gamma_M|$ denotes
the volume of $\Gamma_M$.} The microstate of a system inherits the
entropy and information of the macrostate to which it belongs. The
entropy of an isolated system increases because the microstate of
the system moves into macrostates of ever greater entropy. The
equation above means that the information possessed by a system at
a point in time is the difference between the maximum entropy of
the system, and the entropy possessed by the system at that point
in time. The maximum information which can be possessed by a
system is that which it possesses when the system's entropy is
zero. Hence, according to the relationship above, the maximum
information equals the maximum entropy.

Whether this entails that a finite volume of space possesses a
finite number of states is a different question. In classical
mechanics, a system consisting of n particles has a 6n-dimensional
\emph{continuum} state space $\Gamma$, called the phase space. The
entropy $S(\rho)$ of a statistical state $\rho$ in classical
mechanics is defined to be

$$
S(\rho) = -k_B \int_\Gamma \rho \log \rho \; d\mu \; ,
$$ where $k_B$ is Boltzmann's constant. In the case of a macrostate $\rho_M$,
this reduces to

$$ \eqalign{
S(\rho_M) &= - k_B \int_{\Gamma_M} |\Gamma_M|^{-1} \log
|\Gamma_M|^{-1} d \mu \cr &= k_B \log |\Gamma_M|\; . } $$

Hence, although the entropy of a macrostate of such a system can
be finite, it corresponds to a continuum of possible microstates.
An upper limit to entropy does not entail a finite number of
possible states. I propose that the link between entropy and
information storage capacity is only valid for finite state-space
systems. When a system has an infinite number of states, but a
finite maximum entropy, I propose that it has an infinite
information storage capacity. Ultimately, each different state of
a system can represent different information, so a system with an
infinite number of possible states, but a finite volume state
space, and therefore a finite maximum entropy, nevertheless has an
infinite information storage capacity.

To argue that a finite volume of space possesses a finite
information storage capacity, one might alternatively start from
loop quantum gravity, and try to argue that a finite volume of
space only possesses a finite number of quantum states. A finite
volume of space corresponds to a finite number of spin network
nodes, and for a fixed finite number of nodes, there are a finite
number of spin network states. For a system with a finite number
of microstates, each macrostate $M$ corresponds to an equivalence
class containing a finite number of microstates, $\text{Num}(M)$.
The entropy of such a macrostate is simply

$$
S = k_B \; \log \text{Num}(M)\;.
$$ Hence, a system with a finite number of microstates possesses a
finite maximum entropy, and an upper limit on its information
storage capacity.

Quantum theory, however, may not be the definitive theory of the
physical world. A quantum state may correspond to many, or an
infinite number of actual states. Even though there may be only a
finite number of quantum states for a finite volume of space,
there may be an infinite number of actual states. It may be that
quantum theory is only valid for certain levels of structure, and
it might merely be that the amount of information which can be
coded above a certain length scale, or the amount of information
which can be coded in a certain way, is finite.

There is also no decisive evidence that elementary particles
exist. If the current candidates for elementary particles, such as
quarks, do have parts, then those parts might only be detectable
at energies which are not currently available in particle
accelerators.

One could also dispute the assumption that, on each level of
structure, there is a finite set of parts in any finite volume of
space. If each part has a non-zero spatial extension with a
well-defined boundary, and if the parts cannot inter-penetrate,
then it does indeed follow that there can only be a finite set of
parts packed into a finite volume of space. However, parts in
quantum theory do seem able to interpenetrate each other to some
degree. If there are levels of structure below the levels of the
electron and quark, these might reveal very strange things, beyond
the imagination even of quantum theory, such as an infinite number
of parts interpenetrating each other in a finite volume of space.

\hfill \break

Tipler claims that there could be a hierarchy of computer
universes, just like the hierarchy of so-called `virtual machines'
which can exist on a computer, and he claims that we would not
know which level of the hierarchy our own universe exists at.
Whilst I have argued that the Bekenstein bound does not entail
that a finite volume subsystem has only a finite number of
possible states, Tipler accepts this implication. This, I propose,
is inconsistent with the claim that we would not know which level
of a universe hierarchy our own universe exists at.

When one computer is programmed so that it precisely mimics the
input-output behaviour of another computer, the latter is said to
be emulated by the former. The emulation program, running on the
real computer, is said to be a virtual machine. A real machine
$T_1$ can be programmed to emulate another, producing a virtual
machine $T_2$. The virtual machine $T_2$ can then be programmed to
emulate another computer, producing a higher level virtual machine
$T_3$. These levels are referred to as levels of implementation.

A universe running on a computer could itself contain computers,
upon which other universes are running. The universes would be
running at different levels of implementation, and Tipler
suggests, (1995, p208), that in this case, the levels should be
thought of as levels of reality. Tipler seems to assume that there
must be a lowest level of the hierarchy, and refers to this as
`ultimate reality'. He claims that ``we cannot know if the
universe in which we find ourselves is actually ultimate reality,"
(ibid.).

However, whilst any one computer may be able to emulate the
input-output behaviour of another, that does not entail that any
one computer has the same representational capacity as another. An
actual computer, with a finite memory, does not have the same
representational capacity as every other computer. A computer with
$N$ bytes of memory does not have the same representational
capacity as a computer with $M$ bytes of memory if $M > N$. There
may be data structures which the computer with $M$ bytes of memory
can represent, but which the computer with $N$ bytes cannot.

It was argued above that a computer with a finite set of states,
(and hence a finite memory), cannot perfectly represent the
universe to which it belongs. This is because a computer with a
finite memory cannot code the same amount of information as the
universe to which it belongs. In general, a computer with a finite
memory cannot perfectly represent any universe which can code a
greater amount of information than the computer. Any universe
which can code a greater amount of information than the universe
to which the computer belongs, will code more information than the
computer.

If one accepts Tipler's claim that ``complexity is appropriately
measured by the number of possible alternative states a system can
be in," (1995, p118), then the complexity of a system can also be
measured as the amount of information which that system can
code.\footnote{This should not be confused with the
\emph{computational complexity} of an algorithm used to calculate
the values of a function. This is a measure of the growth in
computation time with the growth of the size of the input. For
example, those functions which are computable by an algorithm in
polynomial time, are referred to as \textbf{P} problems. (See
Penrose (1989), p181-187, for a good introduction). Tipler's
notion of complexity is also distinct from the \emph{Kolmogorov
complexity} of an object, also known as the \emph{algorithmic
complexity}. The Kolmogorov complexity of a bit-string is the
length, in bits, of the shortest computer program capable of
producing that bit-string as output. By extension, if one has a
digital representation of an object by a bit-string, one can
define the Kolmogorov complexity of that representation to be the
length, in bits, of the shortest computer program capable of
producing that digital representation as output.} If one accepts
that a finite volume subsystem has only a finite number of
possible states, then a computer can only have a finite memory. If
a computer can only have a finite memory, then a computer cannot
perfectly represent a universe of the same complexity, or greater
complexity, than the universe to which the computer belongs. The
complexity of a universe is observable, hence, \textit{contra}
Tipler, the levels of implementation are distinguishable. If a
finite volume subsystem has only a finite number of possible
states, then each higher level of universe implementation is less
complex than the level below. A computer with a finite memory
cannot perfectly represent a universe unless that universe is
simpler than the universe to which the computer belongs. The more
complex the universe one belongs to, the lower down the hierarchy
that universe is placed. A universe of maximal complexity, if
there is such a thing, could be proven to be the universe of
ultimate reality.

If our universe is a computer program running on a computer in
another universe, then that universe must have a higher level of
complexity to our own. This greater complexity might take the form
of a higher number of spatial dimensions.

Of course, if a finite volume subsystem has a discrete infinity of
possible states, then a computer might be able to perfectly
represent a universe with the same complexity as the universe to
which the computer belongs. If so, then the levels of universe
implementation might all have the same level of complexity. The
point is that, if the Bekenstein bound does entail that a finite
volume subsystem has only a finite number of possible states, then
the Bekenstein bound is inconsistent with the thesis that
universes at different levels of implementation are
indistinguishable.

\section{Supervenience, identity, and universe creation on a computer}

The suggestion that a physical system can be perfectly simulated
on a computer is consistent with the principle of supervenience,
but suggests that a physical system can be realised on more than
one medium. Suppose, for example, that a tornado could be
perfectly simulated on a computer. A tornado is described by a
solution of the Navier-Stokes equations.\footnote{There is, for
example, an exact solution of the Navier-Stokes equations called
the Sullivan Vortex, which describes the flow in an intense
tornado with a central downdraft.} To simulate a tornado on a
computer, one would define program variables to represent the air
pressure, velocity, density etc. in a volume of space, and one
would represent the tornado by calculating a solution of the
Navier-Stokes equations for these variables. Whilst a `real'
tornado is a process running on a collection of air molecules, a
simulated tornado is a process running upon the components and
circuitry of a computer. Hence, if a tornado could be perfectly
simulated on a computer, one might argue that a tornado could be
realised upon more than one medium. The processes associated with
two completely different lower-level media, appear to be capable
of yielding the same higher-level process. In the particular case
of the simulation of a mind on a computer, Bostrom describes this
as the notion of `substrate-independence', arguing that ``mental
states can supervene on any of a broad class of physical
substrates," (Bostrom 2003, Section II).

Supervenience basically proposes that the parts of a system, and
the way in which the parts are organized and interact, uniquely
determine the higher-level states and properties of the system. In
other words, the states and properties of the subsystems in a
composite system, and the relationships between the subsystems,
uniquely determine the higher-level states and properties of the
composite system. The idea is that there can be no difference in
the higher-level state of a composite system without a difference
in the lower-level state, otherwise one would have a one-many
correspondence between the lower-level states and higher-level
states.

If a physical system could be realised on more than one medium, it
would not undermine the principle of supervenience. For example,
the properties of a tornado might not determine a unique medium
upon which it must be realised, but the properties of air
molecules, and the relationships between air molecules, entail
that a tornado can be realised on a collection of air molecules.
Similarly, if it were possible to realise a tornado on a computer,
then it would be the properties of, and relationships between, the
components and circuitry of a computer which would entail that a
tornado could be realised upon a computer.

Whether or not a physical system can in fact be realised on more
than one medium depends upon how one defines the identity of a
system. In the case of a tornado there are two possible
approaches:

\begin{enumerate}
\def\theenumi{{\rm(\alph{enumi})}}
\item{A tornado is a physical system composed of atmospheric
molecules, which has the property that it satisfies a
tornado-solution of the Navier-Stokes equations. The identity of a
tornado is inseparable from being a collection of atmospheric
molecules. A tornado is not as much realised upon a collection of
atmospheric molecules, as it is composed of atmospheric molecules.
A tornado cannot be realised on more than one medium because there
is no sense in which a tornado is realised on any medium. It is
only if the identity of a tornado could be defined in a formal,
mathematical sense, that one could speak of a tornado being
realised upon a medium.}

\item{The identity of a tornado is defined by a solution of
the Navier-Stokes equations, and the identity of a solution of the
Navier-Stokes equations is independent of any particular medium,
hence the identity of a tornado is independent of any particular
medium. The identity of a tornado is independent of its
realisation upon a collection of atmospheric molecules. If the
components and circuitry of a computer could realise a
tornado-solution of the Navier-Stokes equations, then a tornado
could be realised on the components and circuitry of a computer.}
\end{enumerate}

The identity of a solution to the Navier-Stokes equations is
independent of any particular physical medium because a solution
of a differential equation is merely a mathematical object. A
solution to a differential equation is given physical meaning when
the solution variables are given a physical reference i.e.
physical units. The solution variables of a differential equation
can refer to many different things: consider, for example, the
diverse domains in which one can find solutions to the wave
equation or the diffusion equation.

When a solution of the Navier-Stokes equations is realised on a
medium, the solution variables have physical referents. When a
solution of the Navier-Stokes equations is realised on the medium
of atmospheric molecules, the solution variables refer to air
pressure, velocity, density etc. If, alternatively, a solution
could be realised on, say, an economic system, then the solution
variables would refer to economic quantities instead.

For a computer to be able to realise a tornado-solution of the
Navier-Stokes equations, the computer must possess objective
physical properties which could be the referents of the
tornado-solution variables. Whilst it is permissable for these
properties of the computer to be compound or collective
properties, they must be objective physical properties. If a
tornado-solution were to be realised on a computer, the solution
variables would not refer to properties of the atmosphere, such as
pressure, velocity, density etc. Instead, they would refer to
properties of the computer components and circuitry, such as,
perhaps, the voltage states of the bytes in computer memory. To
reiterate, the medium upon which a solution is realised is defined
by the referents assigned to the solution variables.

It is possible to accept approach (b), that the identity of a
tornado is independent of any particular medium, without accepting
that a tornado can be realised on a computer. A computer does not
possess objective physical properties which can be the referents
of the solution variables for the Navier-Stokes equations. One
reason is that the solution variables are continuous, whilst the
logical states of electronic circuits are discrete. A
tornado-solution to the Navier-Stokes equations is probably a bad
example at this juncture because the Navier-Stokes equations, and
fluid mechanics \textit{in toto}, merely provide a
phenomenological approximation.\footnote{i.e Fluid mechanics is
able to explain and predict a range of macroscopic phenomena to a
certain degree of approximation, but more fundamental theories are
required to describe what actually exists and happens.} A
tornado-solution of the Navier-Stokes equations is not exactly
realised on the medium of air molecules either. However, even if
one goes down to the level of fundamental physics, a computer
cannot exactly realise solutions to the fundamental equations of
physics either. The reason is twofold:

\begin{itemize}
\item{There is a one-many correspondence between the logical states
and the exact electronic states of circuits.} \item{The logical
states of multiple bits in computer memory only represent numbers
because they are deemed to do so under a numeric-interpretation.}
\end{itemize}

In contemporary digital computers, each bit of memory corresponds
to an electrical circuit, and the two possible logical states of
the bit correspond to different possible voltages between fixed
points of the circuit. The logical state of $1$ is not defined by
a single precise voltage value, but by a range of values, and the
logical state of $0$ is defined by a different range of possible
voltages. There is, therefore, a one-many correspondence between
logical states and voltage levels. Successive runs of the same
program will not produce exactly the same sequence of electronic
states in computer memory. The exact voltage levels will be
different on successive runs.

This level of electrical noise prevents a contemporary digital
computer from exactly realising anything, even discrete objects.
Given the one-many correspondence between logical states and exact
electronic states, the exact electronic properties of a computer's
components cannot be the referents of the Navier-Stokes solution
variables. At best, this suggests that a tornado-solution of the
Navier-Stokes equations could only be approximately realised on a
digital computer. This is crucial to the question of whether the
same physical system can be realised on more than one physical
medium. If there cannot be an exact realisation of a tornado on
the medium provided by the components and circuitry of a computer,
this is presumably because the properties of, and relationships
between, the components and circuitry of a computer differ from
the properties of, and relationships between, the air molecules in
a region of the atmosphere.

Moreover, it is not even possible to contend that a
tornado-solution of the Navier-Stokes equations could be
approximately realised on a digital computer. A computer
simulation provides no type of realisation at all. It is not the
logical states of multiple electrical circuits in computer memory,
but the \emph{numeric interpretation} of the logical states which
are the candidates to be referents of the tornado solution
variables. It is the pattern of numbers represented by a computer
which resembles the pattern of values realised by a simulated
system's physical quantities. As explained in the next section,
the numbers represented by a computer are
interpretation-dependent, hence the numbers represented by a
computer cannot be objective physical properties of the computer.
If the numbers represented by the computer are
interpretation-dependent, then the pattern of numbers represented
by the computer must be an interpretation-dependent pattern.
Hence, the resemblance between the pattern of numbers represented
by the computer and the pattern of values possessed by the
physical quantities of a simulated system, must be an
interpretation-dependent resemblance. Change the interpretation of
the logical states of the multiple electrical circuits in computer
memory, and there is no resemblance, not even an approximate one.
Even if there was no electrical noise, and even if the simulated
system was discrete itself, (even if there was a bijective
correspondence), it would still be an interpretation-dependent
resemblance. The numbers represented by the computer are not
objective physical properties of the computer. To constitute a
realisation of a Navier-Stokes tornado-solution, the referents of
the solution variables must be objective physical properties, not
interpretation-dependent, hence a computer cannot realise a
tornado-solution of the Navier-Stokes equations, or any other
physical system for that matter.

\section{A digital computer simulation of a universe cannot exist as a universe}

A digital computer simulation of a physical system cannot exist
as, (does not possess the properties and relationships of),
anything else other than a physical process occurring upon the
components of a computer. In the contemporary case of an
\emph{electronic} digital computer, a simulation cannot exist as
anything else other than an electronic physical process occurring
upon the components and circuitry of a computer. The following
argument will be deployed to establish this conclusion:

\begin{enumerate}
\item{A digital computer simulation is a type of representation.}
\item{There are three types of representation.}
\item{A digital computer simulation is a special case of the
type of representation in which there is no objective
relationship, and in particular no homomorphy, between the
represented thing and the thing which represents it.}
\item{If there is no objective relationship between a universe
and a digital computer simulation of a universe, then a digital
computer simulation of a universe cannot exist as a universe.}
\end{enumerate}

The reasoning that justifies claim 3, outlined at the end of the
previous section, is basically as follows: In a computer
simulation, the values of the physical quantities possessed by the
simulated system are represented by the combined states of
multiple bits\footnote{Qubits in the case of quantum computers.}
in computer memory. However, the combined states of multiple bits
in computer memory only represent numbers because they are deemed
to do so under a numeric interpretation. There are many different
interpretations of the combined states of multiple bits in
computer memory. If the numbers represented by a digital computer
are interpretation-dependent, they cannot be objective physical
properties. Hence, there can be no objective relationship between
the changing pattern of multiple bit-states in computer memory,
and the changing pattern of quantity-values of a simulated
physical system.

Because a digital computer simulation of a universe cannot exist
as a universe, it is, \textit{a fortiori}, impossible for anyone
to be embedded in a digital computer simulation. It is impossible
for our experience to be indistinguishable from the experience of
someone embedded in a digital computer simulation because it is
impossible for anyone to be embedded in a digital computer
simulation.

\hfill \break

Tipler and Bostrom both assume that if a universe is simulated on
a computer, then the simulation exists as a universe, at a
so-called `higher level of implementation'. This ontological
assumption can be generalized to the following proposition: If a
physical system of type $\mathscr{T}$ is simulated on a computer,
then the simulation exists as a system of type $\mathscr{T}$, at a
higher level of implementation. For example, if a tornado is
simulated on a computer, it could be claimed that the simulation
exists as a tornado, at a higher level of implementation. In
opposition, it will be argued in this section that a digital
computer simulation of a physical system, even a perfect
simulation, cannot exist as the thing it represents.

A computer simulation is a special type of representation. In
general, a representation is defined by a mapping $f$ which
specifies the correspondence between the represented thing and the
thing which represents it. An object, or the state of an object,
can be represented in two different ways:

\begin{enumerate}
\item{If an object/state is a structured entity $M$, it can provide the
entire domain of a mapping $f: M \rightarrow f(M)$ which defines
the representation. The range of the mapping, $f(M)$, is also a
structured entity. The mapping $f$ is a homomorphism with respect
to some level of structure possessed by $M$ and $f(M)$.}
\item{An object/state can be an element $x \in M$ in the domain of
a mapping $f: M \rightarrow f(M)$ which defines the
representation.}
\end{enumerate}

The representation of a Formula One car by a wind-tunnel model is
an example of type-1 representation. There is an approximate
homothetic isomorphism\footnote{A transformation which changes
only the scale factor.} from the exterior surface of the model to
the exterior surface of a Formula One car. This notion of
structure preservation can be seen in other cases of
representation. The notorious map of the London Underground does
not preserve geometry, but it does preserve the topology of the
network. Hence in this case, there is a homeomorphic isomorphism
involved.

Type-2 representation has two sub-types. The mapping $f: M
\rightarrow f(M)$ can be defined by either (2i) an objective,
causal physical process, or by (2ii) the decisions of
thinking-beings. The three different types of representation are
similar to C.S. Peirce's tripartite division of representational
`signs' into `icons', `indices', and `symbols'. Peirce held that
icons resemble what they represent, indices are causally connected
to what they represent, and symbols are arbitrary labels for what
they represent, (see Schwartz 1995, p536-537).

The primary example of type-2i representation is the
representation of the external world by brain states. Taking the
example of visual perception, there is no homomorphism between the
spatial geometry of an individual's visual field, and the state of
the neuronal network in that part of the brain which deals with
vision. However, the correspondence between brain states and the
external world is not an arbitrary mapping. It is a correspondence
defined by a causal physical process involving photons of light,
the human eye, the retina, and the human brain. The correspondence
exists independently of human decision-making.

As an example of type-2ii representation, the state of a light
switch could be used to represent things other than itself. One
could decide that the On-position of a light switch represents the
number $1$, and the Off-position represents the number $0$. This
relationship between the states of the light switch and the set
$\{0,1\}$ does not exist objectively. In other words, the
relationship does not exist independently of the interpretative
decisions made by human-beings. Someone else could decide that the
On-position represents the number $0$, and that the Off-position
represents the number $1$. One could even decide that the
On-position of a light switch represents the colour black, and the
Off-position represents the colour white. There is no homomorphism
between the On-position of a light switch and either the number
$1$ or the colour black. The position of the light switch is
merely being used as an element in the domain of a mapping which
defines the representation.

In the case of a digital computer simulation, the bytes of memory
are used to represent numbers and numbers are used to represent
the quantities of the simulated system. Hence, the representation
of a tornado by the logical states of a current digital computer
is an example of type-2ii representation. There is no homomorphism
between the electronic states or logical states of a current
digital computer and the things those states are chosen to
represent.\footnote{Recall that there is a one-many correspondence
between the logical states and the exact electronic states of
computer memory. Although there are bijective mappings between
numbers and the logical states of computer memory, there are no
bijective mappings between numbers and the exact electronic states
of memory.} The logical states of a computer can be mapped to many
different things, (numbers, images, and sounds etc), but in each
case a logical state is merely an element in the domain of the
mapping which defines the representation. The logical state of a
computer is not the domain of a homomorphic mapping, and human
decisions, rather than causal processes, determine what things the
logical states of a digital computer represent. For these reasons,
the states of a digital computer are not objectively related to
that which they are deemed to represent.

Whilst the electronic states of a current digital computer do
indeed possess a quite intricate structure, that structure is not
used for the representational applications of a computer. The
state of each bit in the memory of a computer is defined by the
1-dimensional graph topology of an electrical circuit, and by the
voltage between specific points of the circuit. Hence, the
memory-state of a computer is something which possesses a quite
intricate structure. However, this electrical circuit and voltage
structure bears no resemblance to the things which the memory of a
computer is deemed to represent.

If a digital computer simulation of a universe is a type-2ii
representation, then a digital computer simulation of a universe
is not objectively related to that universe. This rules out the
claim that a digital computer simulation could exist as a
universe.

Bostrom's hypothesis that we could be living in a computer
simulation assumes that ``it would suffice for the generation of
subjective experience [in a computer simulation] that the
computational processes of a human brain are structurally
replicated in suitably fine-grained detail," (Bostrom 2003,
Section II). The notion of `structural replication' is the same as
the notion of isomorphism, so Bostrom's hypothesis that we could
be living in a computer simulation is based upon the false
assumption that a computer simulation provides the type of
representation in which an isomorphism or homomorphism exists.
Hence, even if one endorses the notion of `structural realism',
that a thing is completely defined by its structural mathematical
relationships, one cannot say that a digital computer simulation
realises the thing it represents.

\hfill \break

Although the states of a \emph{digital} computer are not
objectively related to the things they are deemed to represent, it
is possible that the states of an analog computer could be so
related. It is conceivable that there could be a homomorphism
between the states of an analog computer and the things those
states represent. Whilst an analog computer does not necessarily
resemble the system it represents in terms of geometry or
topology, a homomorphism between physical objects is not
necessarily a homomorphism of spatial geometry or topology. The
examples of a wind-tunnel model and the London Underground map are
misleading in this respect. The homomorphism could be a non-visual
homomorphism. An analog computer could possess objective physical
properties which change with the same pattern as the changing
pattern of values for the physical quantities on a simulated
system. Hence, an analog computer simulation might provide type-1
representation.  If this is so, then a more general argument would
be required to demonstrate that no type of computer simulation at
all could exist as a universe.

In the examples of type-1 representation given above, although
there is a physical resemblance in some respects between $M$ and
$f(M)$, there is not a total resemblance. For example, although
the parts of a wind-tunnel model subtend the same angles as the
actual car, the wind-tunnel model is not the same size as the
actual car. Despite such examples, there is no reason in principle
why a type-1 representor cannot possess all the properties of the
thing it represents. At least, there is no reason why a type-1
representor cannot possess all the `intrinsic' properties of the
thing it represents.\footnote{The intrinsic properties of an
object are the properties it possesses independently of its
relationships to other objects. If one object is numerically
distinct from another, as a representor must be from what it
represents, then the two objects cannot share the same set of
relationships with other objects. To be numerically distinct, they
must occupy disjoint regions of space-time, and therefore cannot
share the same set of spatio-temporal relationships with other
objects.}

If a type-1 representor possesses all the intrinsic properties of
the thing it represents, then one might conclude that it exists as
the same type of thing as the thing it represents. Accordingly, an
analog computer simulation of a universe might exist as a
universe. However, to reiterate a point made in section 3, it
remains to be proven that an analog computer can possess the
representational capacity to represent an entire universe.

Tipler and Bostrom both imagine a computer simulation which would
simulate all the people who exist in our own universe. Such
simulated people, it is suggested, would reflect upon the fact
that they think, would interact with their apparent environment,
and would conclude that they exist. The claim that a simulated
universe would be real to the simulated people, presupposes that
simulated systems provide realisations of those systems, and
presupposes that simulated people exist as people. Digital
computer simulations of people exist only as physical processes on
a computer, not as people. Hence, there are no people in a digital
computer simulation to reflect upon the fact that they think, or
to interact with their apparent environment.

If a digital computer simulation of a universe cannot exist as a
universe, then the sceptical hypotheses of Tipler and Bostrom
cannot be true. It is impossible that our own experience is
indistinguishable from the experience of somebody embedded in a
digital computer simulation because it is impossible for anybody
to be embedded in a digital computer simulation. Systems cannot be
realised in digital computer simulations, and people cannot exist
in digital computer simulations.

\end{document}